\documentclass[a4paper]{article}

\usepackage[margin=3cm]{geometry}
\usepackage{amsmath, amsthm, mathtools} 
\usepackage[small]{titlesec}
\usepackage{authblk}
\usepackage[default]{fontsetup} 
\usepackage{microtype} 
\usepackage{tikz-cd}
\usepackage{graphicx}
\usepackage{hyperref}

\theoremstyle{definition}
  \newtheorem{thm}{Theorem}[section]

  \newtheorem*{rem}{Remark}

\newcommand{\scr}{\symscr}

\newcommand{\on}{\operatorname}

\newcommand{\la}{\langle}
\newcommand{\ra}{\rangle}

\newcommand{\R}{\symbb{R}}

\newcommand{\Z}{\symbb{Z}}

\newcommand{\bw}{{\textstyle\bigwedge}}

\newenvironment{blk}[1]
  {\begin{trivlist}
   \item[\hskip \labelsep {\bfseries #1.}]
  }
  {\end{trivlist}}

\title{ \textsc{Forms, half-densities, and the quantum odd symplectic category in the BV formalism}}
\author{Pavol Ševera}
\affil{\small Section of Mathematics\\University of Geneva, Switzerland \\  \texttt{pavol.severa@unige.ch}}
\date{}

\begin{document}
\maketitle

\begin{abstract}
  This note is a detailed review of the geometry behind the Batalin-Vilkovisky formalism and how it fits into the framework of the quantum odd symplectic category and the odd quantization functor. 
\end{abstract}

\section{Introduction}

The discovery by H.~Khudaverdian of  the natural odd Laplacian operator $\Delta$ acting on half-densities on an odd symplectic manifold \cite{kh} gave a geometric explanation of the Batalin-Vilkovisky formalism, completing a work started by A.~Schwarz \cite{sch}. At the same time, it uncovered a previously hidden geometry of differential forms on manifolds, replacing the usual symmetry group $\on{Diff}(M)$ of diffeomorphisms by the larger supergroup $\on{Sympl}(\Pi\, T^* M)$ of symplectomorphisms of the odd cotangent bundle. This new geometry is closely related to the calculus operad of D.~Tamarkin and B.~Tsygan \cite{ta-ts} which appeared soon after Khudaverdian's work; the operad can be seen as the collection of algebraic and differential operations which can be done with half-densities and functions on an odd symplectic manifold.

If $M$ is a manifold then  half-densities on the odd symplectic manifold $\Pi\, T^*M$ can be naturally identified, via the fibrewise Fourier transformation, with functions on $\Pi\, TM$, i.e.\ with differential forms on $M$. Under this identification $\Delta$ is simply the de Rham differential on $\Omega(M)$. Moreover, integration of differential forms over oriented submanifolds of $M$ becomes a special case of an integration of half-densities over oriented Lagrangian submanifolds of $\Pi\, T^* M$. In this way, Khudaverdian's calculus of half-densities becomes a more flexible version of the usual calculus of differential forms.

The link with differential forms is present not just in the geometric formulation of the BV formalism, but also in the actual use of the BV method in perturbative quantum field theory: In the presence of a moduli space,  perturbative calculations don't produce numbers, but rather differential forms on the moduli space. Differential forms show up naturally also if we consider families of QFTs. In all of these cases it is very helpful to see the forms as half-densities on odd symplectic manifolds.

The purpose of this note is twofold. Much of the literature on this subject is written in a somewhat careless way, without spelling out all the assumptions and even all the needed data. This is particularly true for my work \cite{se-qosc} about the odd symplectic category and the odd quantization functor, which subsume and extend the methods used in the BV formalism, including, in particular, the BV fibre integral. I therefore think that a full and clean statement of all the relevant definitions and theorems is needed. This review is done in Sections \ref{sec:half-dens} and  \ref{sec:qosc}. The second (though  minor) purpose is to show the power of these basic ideas by rephrasing, in a simple way, an ingenious construction involving families of Lagrangian submanifolds and differential forms, which appeared in \cite{mi-sch} and which was used in \cite{ge-po}. The rephrasing is done purely in the language of half-densities and Lagrangian submanifolds, which makes it much more transparent, to the point of being tautological. This is done in the final Section \ref{sec:families}.

I'd like to thank Ján Pulmann for persuading me to write this note and for his remarks on its preliminary version.

\section{Half-densities and BV-operator on odd symplectic manifolds}\label{sec:half-dens}

In this section we review Khudaverdian's calculus of half-densities on odd symplectic manifolds introduced in \cite{kh}. For a recent review see \cite{cat}. We shall  use the approach of \cite{se}, since it uses simpler calculations and uncovers curious relations. 


\subsection{The odd Laplacian as a differential}
The first main statement is that if $\scr M$ is an odd symplectic manifold with local Darboux coordinates $x^i$ (even) and $\pi_i$ (odd) and the symplectic form $\omega = dx^i\wedge d\pi_i$, then the operator
\begin{equation}\label{eq:Delta-coord}
  \Delta =\frac{\partial^2}{\partial x^i \partial \pi_i}
\end{equation}
acting on half-densities on $\scr M$ is independent of the coordinates. 

The proof/construction from \cite{se} is as follows. There is a natural square root of the Berezinian on $\scr M$. Its sections, \emph{half-densities} on $\scr M$, are the elements of the cohomology (which is a $C^\infty(\scr M)$-module)
\begin{equation}\label{eq:dens12def}
  \on{Ber}^{1/2}(\scr M) \coloneq H\bigl(\Omega(\scr M), \omega\wedge\bigr)
\end{equation}
($(\Omega(\scr M), \omega\wedge)$ is understood here as a $\Z/2\Z$-graded complex).

In local Darboux coordinates each cohomology class has a unique representative of the form
\begin{equation}\label{eq:representant}
  f(x, \pi)\,dx^1\wedge\dots\wedge dx^n.
\end{equation}
Indeed, a simple calculation shows that with the homotopy operator
\[J = i_{\partial_{x^i}} i_{\partial_{\pi_i}}\]
we get
\begin{equation}\label{eq:n-dg}
  [\omega\wedge, J] = n - \on{sdeg}
\end{equation}
where
\[  \on{sdeg} = dx^i\, i_{\partial_{x^i}} - d\pi_i\, i_{\partial_{\pi_i}}\]
multiplies by the number of $dx$'s minus the number of $d\pi$'s. In particular,   the $\on{sdeg}=n$ eigenspace contains the forms \eqref{eq:representant}. The complex $\bigl(\Omega(\scr M),\omega\wedge\bigr)$ splits into a direct sum of the eigenspaces of $\on{sdeg}$.  Since  \eqref{eq:n-dg} implies all these eigenspaces are acyclic except for  $\on{sdeg}=n$  (where the differential vanishes), every class has a unique representative of the form \eqref{eq:representant}.

The class of a form \eqref{eq:representant} can then be understood as a half-density (written a bit formally)
\[f(x, \pi)\, \sqrt{dx\,d\pi}.\]
In \S\ref{ssec:Lagrangian} we shall explain why these objects are really half-densities, i.e.\ why the product of two half-densities is a  section of the Berezinian bundle,  but the expression \eqref{eq:representant} makes it reasonably clear.

Since $\omega$ is closed, the two operations $\omega\wedge$ and $d$ turn $\Omega(\scr M)$ to a bicomplex. In the corresponding spectral sequence the differential $d$ on $H\bigl(\Omega(\scr M), \omega\wedge\bigr)$ vanishes, since it goes from the $\on{sdeg}=n$ eigenspace to the $\on{sdeg}=n-1$ eigenspace. The next differential is 
\[\Delta = [d\circ (\omega\wedge)^{-1} \circ d] \colon H\bigl(\Omega(\scr M), \omega\wedge\bigr) \to H\bigl(\Omega(\scr M), \omega\wedge\bigr),\]
that is, if $\alpha\in\Omega(\scr M)$ satisfies $\omega\wedge\alpha =0$ and if $\beta\in\Omega(\scr M)$ is such that $d\alpha=\omega\wedge\beta$, then
\[\Delta [\alpha] = [d\beta].\]
We have, as we wanted,
\begin{equation}\label{eq:Delta-rep-coord}
  \Delta\bigl( f(x, \pi)\,[dx^1\wedge\dots\wedge dx^n]\bigr) = \frac{\partial^2f}{\partial x^i \partial \pi_i}\,[dx^1\wedge\dots\wedge dx^n].
\end{equation}
Indeed,
\begin{align*}
  f(x, \pi)\,dx^1\wedge\dots\wedge dx^n &\xrightarrow{d} \frac{\partial f}{\partial \pi_i}\,d\pi_i\wedge dx^1\wedge\dots\wedge dx^n
  \xrightarrow{J} i_{\partial_{x^i}}\,\frac{\partial f}{\partial \pi_i}\, dx^1\wedge\dots\wedge dx^n\\
  &\xrightarrow{d} \frac{\partial^2f}{\partial x^i \partial \pi_i}\,dx^1\wedge\dots\wedge dx^n + \dots
\end{align*}
where the  term  $\dots$ has $\on{sdeg} = n-2$ and therefore is $\omega\wedge$-exact.

\begin{rem}[Homotopy transfer]
  The above construction can be understood as a quasi-isomorphism of the complexes $\bigl(\Omega(\scr M), {\omega\wedge} + t\,d\bigr)$ and $\bigl(\on{Ber}^{1/2}(\scr M),-t^2\Delta\bigr)$ (where $t$ is an auxiliary formal variable) obtained via the homological perturbation lemma (HPL, see \cite{cr} for a review). Indeed, we have the quasi-isomorphisms (an embedding and a projection) of complexes
  \[\bigl(\on{Ber}^{1/2}(\scr M),0\bigr) \xrightleftharpoons[p]{\iota} \bigl(\Omega(\scr M), {\omega\wedge}\bigr)\]
  where we see $\on{Ber}^{1/2}(\scr M)$ as the $\on{sdeg}=n$ eigenspace and where $p$ projects out the other eigenspaces. If we define a homotopy operator $\hat J$ on $\Omega(\scr M)$ to be on the $\on{sdeg}=k$ eigenspace
  \[\hat J = 
    \begin{cases}
      -(n-k)^{-1}J & \text{if $k\neq n$}\\
      0 & \text{if $k=n$}
    \end{cases}
  \]
  then $(\iota,p,\hat J)$ is a special deformation retract (as $J^2=0$ and hence $\hat J^2=0$). HPL then gives us a quasi-isomorphism 
  \[\bigl(\Omega(\scr M), {\omega\wedge} + t\,d\bigr) \simeq \bigl(\on{Ber}^{1/2}(\scr M),\delta\bigr)\]
  where
  \[\delta = p\,d\,\sum_{k=0}^\infty t^{k+1}(\hat J\,d)^k\,\iota.\]
  Since $\hat J$ lowers the degree by 2 and $d$ increases it by 1, only the term with $k=1$ is non-zero, so $\delta=-t^2\Delta$. This is the point of view taken in \cite{abf} and (with minor modifications) in \cite{saf}.
\end{rem}

\begin{blk}{Parity}
  We declare the parity of a half-density $f(x,\pi)\,[dx^1\wedge\dots\wedge dx^n]$ to be the parity of $f(x,\pi)$, i.e.\ we shift the parity by the parity of $n$.
\end{blk}

\begin{blk}{Twisted half-densities}
  If $\Lambda$ is a local system over $\scr M$ (i.e.\ a local system over the body of $\scr M$), it will be convenient to consider  the $\Lambda$-valued half-densities
  \[ \on{Ber}^{1/2}(\scr M, \Lambda) = H\bigl(\Omega(\scr M)\otimes\Lambda, \omega\wedge\bigr) \]
  By the same construction as above we obtain a differential $\Delta$ on  $\on{Ber}^{1/2}(\scr M, \Lambda)$.
\end{blk}

\begin{rem}
  Starting with the works of Schwarz and Khudaverdian, the discussion of $\Delta$ often uses the fact that any $C^\infty$ odd symplectic supermanifold $\scr M$ is non-canonically isomorphic to $\Pi\, T^* M$, where $M$ is the body of $\scr M$. This, however, is not true if we consider families of odd symplectic supermanifolds (the moduli space is $\Pi\, H^2(M;\R)$) or if we go from $C^\infty$ to other categories, so we shall avoid such constructions. Nonetheless, $\Pi\, T^* M$ will play an important role in what follows.
\end{rem}

\subsection{Hamiltonians as homotopies}
The next statement is a form of Cartan's magic formula, saying that Hamiltonian vector fields act on $\on{Ber}^{1/2}(\scr M)$ as $\Delta$-exact operators: for any $h\in C^\infty(\scr M)$ and any  $\sigma\in \on{Ber}^{1/2}(\scr M)$ we have
\begin{equation}\label{eq:CMF}
  [h, \Delta]\,\sigma = - L_{X_h}\sigma
\end{equation}
where $X_h$ is the Hamiltonian vector field generated by $h$. 

To prove it, let us suppose that $h$ is odd (without loss of generality, since if $h$ is even, we simply replace it with $\epsilon h$ where $\epsilon$ is an odd parameter). On $\Omega(\scr M)$ we can summarize the needed identities (using an auxiliary parameter $t$) as
\[ [{\omega\wedge} + t\,d, h + t\,i_{X_h} ] = t^2L_{X_h}. \]
If $\alpha\in\Omega(\scr M)$ satisfies $\omega\wedge\alpha=0$, and so we have a cohomology class $[\alpha]\in\on{Ber}^{1/2}(\scr M)$, then by definition 
\[\Delta[\alpha] = [d\beta]\quad\text{where}\quad d\alpha = \omega\wedge\beta.\]
Since 
\[d(h\,\alpha) = \omega\wedge(h\,\beta - i_{X_h}\alpha), \]
we get
\begin{align*}
  [h,\Delta] [\alpha] &= h\Delta[\alpha] + \Delta[h\,\alpha] = [h\,d\beta] + [d(h\,\beta - i_{X_h}\alpha)] = [dh\wedge\beta - di_{X_h}\alpha]\\
  &= [-i_{X_h}(\omega\wedge\beta) - di_{X_h}\alpha] = -[i_{X_h}d\alpha + di_{X_h}\alpha] = - [L_{X_h}\alpha] = - L_{X_h}[\alpha].
\end{align*}

\subsection{Duality}\label{ssec:duality}

\begin{blk}{Orientation local system}
  For any supermanifold  $\scr S$ of dimension $m|n$ we denote by $\on{Or}_{\scr S}$ the orientation local system of its body,  and we set the parity of $\on{Or}_{\scr S}$ to be the parity of $n$. With this choice the map
  \[ {\textstyle\int}\colon \on{Ber}_c(\scr S, \on{Or}_{\scr S}) \to \R \]
  (where $c$ stands for compactly supported)
  is well-defined and even.
\end{blk}

If $\sigma\in \on{Ber}^{1/2}(\scr M)$ and $\tau\in \on{Ber}^{1/2}_c(\scr M, \on{Or}_{\!\scr M})$
then (as explained below in \S\ref{ssec:Lagrangian}) $\sigma\tau\in \on{Ber}_c(\scr M, \on{Or}_{\!\scr M})$, so we have a well-defined pairing
\[ \la\sigma,\tau\ra \coloneq \int_{\scr M}\sigma\tau.\]
In local Darboux coordinates, if 
\[\sigma = f(x,\pi)\,[dx^1\wedge\dots\wedge dx^n],\quad \tau = g(x,\pi)\,[dx^1\wedge\dots\wedge dx^n],\]
 $\la\sigma,\tau\ra$ is the Berezin integral of $f(x,\pi)g(x,\pi)$.

The coordinate formula \eqref{eq:Delta-coord} for $\Delta$ shows that it is formally selfadjoint, namely that
\begin{equation}\label{eq:Delta-selfadj}
  \la \Delta\sigma, \tau\ra  = (-1)^{|\sigma|}  \la \sigma, \Delta\tau\ra.
\end{equation}

We can now define the space  
\[\on{Ber}^{1/2}_{-\infty}(\scr M)\]
of generalized (i.e.\ distributional) half-densities  as the continuous dual of $\on{Ber}^{1/2}_c(\scr M, \on{Or}_{\!\scr M})$.  $\Delta$ is defined on $\on{Ber}^{1/2}_{-\infty}(\scr M)$  via the duality relation \eqref{eq:Delta-selfadj}. 

\subsection{Lagrangian submanifolds and integration}\label{ssec:Lagrangian}
If $\scr L\subset\scr M$ is an immersed Lagrangian submanifold, half-densities on $\scr M$ restrict in a natural way to sections of the Berezinian bundle on $\scr L$.
Given our definition of half-densities \eqref{eq:dens12def},
the cleanest way of understanding this  is via Manin's cohomological definition of the Berezinian \cite{ma}: If $V$ is a finite-dimensional vector superspace and $W$ a vector superspace with an odd symplectic form $\omega_W$ s.t.\ $V\subset W$ is a Lagrangian subspace (e.g.\ $W=V\oplus \Pi\,V^*$), then
\[\on{Ber}V^* = H\bigl({\textstyle\bigwedge}W^*, \omega_W\wedge\bigr).\]
We then apply the vector-bundle version of this for $V=T\scr L$ and $W= (T\scr M)|_{\scr L}$.

Now we also see why the product of two half-densities is a section of the Berezinian bundle: we just use the Lagrangian submanifold $\scr M_\mathit{diag}\subset\scr M \times \overbar{\scr M}$.

From now on we shall suppose that $\scr L\subset\scr M$ is a properly immersed Lagrangian submanifold.
Supposing that the bodies of $\scr L$ and $\scr M$ are oriented (i.e.\ that $\on{Or}_{\scr L}$ and $\on{Or}_{\!\scr M}$ are trivialized), we can define a generalized  half-density 
\begin{gather*}
  \delta_{\scr L}\in\on{Ber}^{1/2}_{-\infty}(\scr M) \\
  \shortintertext{via}
  \la \delta_{\scr L}, \sigma \ra \coloneq \int_{\scr L}\sigma
\end{gather*}
for any compactly supported half-density $\sigma$ (the integral is well-defined since $\scr L\subset\scr M$ is properly immersed). If  $\scr L$ is given in local Darboux coordinates $x^i$, $\pi_i$ by
\[ \pi_1=\dots =\pi_k=0,\quad x^{k+1}=\dots=x^n=0 \]
then 
\begin{equation}\label{eq:deltaL-coord}
  \delta_{\scr L} = \pm\prod_{i=1}^k\delta(\pi_i)\prod_{\mathclap{j=k+1}}^n\delta(x^j)\,[dx^1\wedge\dots\wedge dx^n]
\end{equation} %
where $\delta(\pi_i)\coloneqq\pi_i$ and the sign depends on the orientation, showing (via \eqref{eq:Delta-rep-coord}) that 
\[\Delta \delta_{\scr L} = 0,\]
which is, to much extent, the basis of the BV method.
In view of \eqref{eq:Delta-selfadj}, it means that
\[\int_{\scr L}\!\Delta\sigma =0 \]
for any compactly-supported $\sigma$. The parity of $\delta_{\scr L}$ is  the parity of $k$.

Let us observe that for any $f\in C^\infty(\scr L)$ we have
\begin{equation}\label{eq:Delta-f-delta}
  \Delta(f\,\delta_{\scr L}) = 0 \quad \Rightarrow \quad f \text{ is locally constant.}
\end{equation}
This follows directly from the coordinate expression \eqref{eq:deltaL-coord}, where $f$ would appear as a function of $x^1,\dots,x^k$ and $\pi_{k+1},\dots,\pi_n$.

In the general (non-oriented) case we have
\begin{equation} \label{eq:where-is-deltaL}
  \delta_{\scr L} \in \on{Ber}^{1/2}\bigl(\scr M, \on{Or}_{\!\scr M} \on{Or}_{\scr L}^{-1}\bigr)^\textit{even}. 
\end{equation}
The local system is defined only in a neighbourhood of $\scr L$, but it's valid since $\delta_{\scr L}$ is supported only on (the body of) $\scr L$.

\subsection{Odd Fourier transform and  differential forms}\label{ssec:odd-Fourier}
If  $M$ is a manifold of dimension $n$, there is a natural isomorphism of ($\Z/2\Z$-graded) complexes, discovered by Khudaverdian,
\begin{equation}\label{eq:hdens-iso-forms}
  \bigl(\on{Ber}^{1/2}(\Pi\,T^*M),\Delta\bigr) \cong \bigl(\Pi^n\,\Omega(M),d\bigr)
\end{equation}
($\Pi^n$ simply means that we change the parity if $n$ is odd and do nothing if $n$ is even).
It is given by the odd Fourier transform along the fibres of $\Pi\,T^*M$. Indeed, in the coordinate expression \eqref{eq:Delta-coord} for $\Delta$, if we apply the odd Fourier transform to the $\pi_i$ variables,  $\Delta$ changes to 
\[d = \xi^i\frac\partial{\partial x^i}\]
where $\xi^i$'s, the  coordinates conjugate to $\pi_i$'s, are then interpreted as $dx^i$'s.

We can describe this isomorphism as follows. If $\scr M = \Pi\,T^*M$ and therefore $C^\infty(\scr M) = \Gamma(\bigwedge TM)$ is the algebra of polyvector fields, the unique local representative \eqref{eq:representant} can be rephrased globally as an isomorphism
\[ \on{Ber}^{1/2}(\Pi\,T^*M) \cong \Gamma\bigl(\bw TM \otimes \bw^{\!\mathrm{top}}T^*M\bigr)\]
and then we simply use the obvious isomorphism of $\on{Cliff}(TM\oplus T^*M)$-modules
\[\bw TM \otimes \bw^{\!\mathrm{top}}T^*M \cong \bw T^*M\]
to get the isomorphism with differential forms.

\subsection{$\Z$-grading}

In applications of the BV formalism the odd symplectic manifold $\scr M$ is typically required to be $\Z$-graded and $\omega$ is required to have degree $-1$. The simplest geometric (if too general) description of this is that we have an even vector field $E$ on $\scr M$  and we require 
\[L_E\omega = -\omega.\]
A function $f\in C^\infty(\scr M)$ is said to be of degree $k$ if $Ef = kf$. Similarly, a half-density $\sigma$ is of degree $k$ if $L_E\sigma = k\sigma$. The operator $\Delta$ is of degree 1.

In this setup $\Pi\, T^*M$ is replaced by its $\Z$-graded version $T^*[-1]M$. The isomorphism \eqref{eq:hdens-iso-forms} becomes the isomorphism of ($\Z$-graded) complexes
\begin{equation}\label{eq:hdens-iso-forms-gr}
 \bigl(\on{Ber}^{1/2}(T^*[-1]M),\Delta\bigr) \cong \bigl(\Omega(M)[n],d\bigr)
\end{equation}
where $n=\dim M$.

\section{Quantum odd symplectic category and the odd quantization functor}\label{sec:qosc}

\subsection{The odd symplectic `category'}

The \emph{symplectic `category'} (including the quotation-mark terminology) was introduced by Alan Weinstein \cite{wein}. Let us spell out its odd version. The objects of the \emph{odd symplectic `category'} $\mathsf{OCS}$ are the odd symplectic manifolds. The morphisms $\scr M_1\to\scr M_2$ are properly immersed Lagrangian submanifolds $\scr L\subset \scr M_2 \times \overbar{\scr M}_1$. Two such Lagrangian relations 
\[\scr L\subset \scr M_3 \times \overbar{\scr M}_2 \qquad \scr L'\subset \scr M_2 \times \overbar{\scr M}_1\]
are \emph{composable} if the projections $\scr L\to \scr M_2$ and $\scr L'\to \scr M_2$ are transverse to each other and if the resulting Lagrangian immersion 
\[\scr L\times_{\scr M_2} \scr L' \to \scr M_3 \times \overbar{\scr M}_1\]
is proper. In that case we set
\[ \scr L\circ \scr L' = \scr L\times_{\scr M_2} \scr L' \colon \scr M_1\to\scr M_3.\]
This `category' is symmetric monoidal, with $\times$ being the monoidal product.
\medskip

\begin{rem}
  The transversality condition is usually weakened to a clean-intersection condition in the ordinary symplectic `category' (and the properness condition typically doesn't appear at all). We shall, however, need both transversality and properness in what follows.
\end{rem}

In general, by a `category' we shall mean the weakening of the notion of category (called also \emph{partial category} or \emph{categoroid}), where not all morphisms can be composed. We do require that $\on{id}_B\circ f = f\circ \on{id}_A$ for any $f\colon A\to B$ (and that both sides are defined), and that if both $f\circ g$ and $g\circ h$ are defined and either $(f\circ g)\circ h$ or $f\circ (g\circ h)$ is defined, then both are defined and
\[(f\circ g)\circ h = f\circ (g\circ h). \]

\subsection{The quantum odd symplectic `category'}

Let us now create a larger `category' out of odd symplectic manifolds, where $\on{Hom}(\scr M_1,\scr M_2)$ is a $\Z/2\Z$-graded complex. It is called \emph{quantum odd symplectic `category'}, $\mathsf{QOSC}$ \cite{se-qosc}. A ``linear'' version of this category was studied in detail in \cite{jpz}.

If $\scr M_1$, $\scr M_2$ are two odd symplectic manifolds, let us define $\on{Hom}(\scr M_1,\scr M_2)$ as the space of continuous linear maps
\[ \on{Ber}^{1/2}_c(\scr M_1) \to \on{Ber}^{1/2}_{-\infty}(\scr M_2). \]
Since both $\on{Ber}^{1/2}_c(\scr M_1)$ and $\on{Ber}^{1/2}_{-\infty}(\scr M_2)$ are $\Z/2\Z$-graded complexes, so is $\on{Hom}(\scr M_1,\scr M_2)$. We shall denote the differential on $\on{Hom}(\scr M_1,\scr M_2)$ simply by $\Delta$.
This is an appropriate notation, since by Schwartz kernel theorem (plus our discussion of $\Delta$) we have an isomorphism of $\Z/2\Z$-graded complexes
\begin{equation}\label{eq:kernels}
  \on{Hom}(\scr M_1,\scr M_2) \cong  \on{Ber}^{1/2}_{-\infty}(\scr M_2\times\overbar{\scr M}_1 , \on{Or}_{\!\scr M_1})
\end{equation}
where $\overbar{\scr M}_1$ is $\scr M_1$ with the opposite symplectic form (this changes the sign of $\Delta$ on $\scr M_1$). In particular, we have
\[ \on{Hom}(*,\scr M) = \on{Ber}^{1/2}_{-\infty}(\scr M),\qquad \on{Hom}(\scr M,*) = \on{Ber}^{1/2}_{-\infty}(\overbar{\scr M}, \on{Or}_{\!\scr M})\]
where $*$ denotes the 1-point manifold. 

The composition of two morphisms $\scr M_1\xrightarrow{F}\scr M_2\xrightarrow{G} \scr M_3$, i.e.\ of
\[ \on{Ber}^{1/2}_c(\scr M_1) \xrightarrow{F} \on{Ber}^{1/2}_{-\infty}(\scr M_2) \quad\text{and}\quad \on{Ber}^{1/2}_c(\scr M_2) \xrightarrow{G} \on{Ber}^{1/2}_{-\infty}(\scr M_3) \]
is defined if $G$ admits a continuous extension to the image of $F$. A sufficient condition for this is given by Hörmander's criteria \cite{ho} using wave-front sets and supports of the kernels. If the compositions are well-defined, we have
\begin{equation} \label{eq:Delta-composition}
  \Delta(F\circ G) = (\Delta F)\circ G + (-1)^{|F|}F \circ (\Delta G).
\end{equation}

Put together, $\mathsf{QOSC}$ is a symmetric monoidal differential $\Z/2\Z$-graded `category'.

\subsection{The oriented odd symplectic `category' and the quantization functor}

\begin{blk}{Orientations of maps}
  If $f\colon\scr S\to \scr T$ is a map between supermanifolds, an \emph{orientation} of $f$ is an isomorphism of principal $\Z/2\Z$-bundles
  \[\nu \colon f^* \on{Or}_{\scr T} \similarrightarrow \on{Or}_{\scr S}\]
  (this is a bit of abuse of notation, since $\on{Or}_{\scr S}$ and $\on{Or}_{\scr T}$ were defined above as the flat vector bundles (local systems) associated to the orientation $\Z/2\Z$-bundles, but hopefully it doesn't cause confusion). We shall say that $\nu$ is even if it preserves the parity and odd if it reverses the parity.
\end{blk}

If $\scr L\subset \scr M_2\times \overbar {\scr M}_1$ is a properly immersed Lagrangian relation, we have, according to \eqref{eq:where-is-deltaL},
\[\delta_{\scr L} \in \on{Ber}^{1/2}\bigl(\scr M_2\times \overbar {\scr M}_1, \on{Or}_{\!\scr M_1}\on{Or}_{\!\scr M_2} \on{Or}_{\scr L}^{-1}\bigr)^\textit{even} \! . \]
To get a $\mathsf{QOSC}$-morphism $\scr M_1\to\scr M_2$ out of $\delta_{\scr L}$ we thus need by \eqref{eq:kernels} an orientation on the projection map $p_2\colon \scr L \to \scr M_2$. We shall denote the resulting $\Delta$-closed morphism in $\mathsf{QOSC}$ by 
\[\delta_{\scr L,\nu}\colon \scr M_1\to\scr M_2.\]
The parity of $\delta_{\scr L,\nu}$ is equal to the parity of $\nu$.

\begin{blk}{Oriented odd symplectic `category'}
  Let us define the \emph{oriented odd symplectic `category'} $\mathsf{OOSC}$ as a slight modification of $\mathsf{OSC}$. Its objects are odd symplectic manifolds  and its morphisms $\scr M_1\to\scr M_2$ are pairs $(\scr L,\nu)$, where $\scr L\subset \scr M_2 \times \overbar{\scr M}_1$ is a properly immersed Lagrangian submanifold and $\nu$ is an orientation of the projection $p_2\colon\scr L \to \scr M_2$. The composition
  \[(\scr L\circ \scr L', \nu\circ\nu') = (\scr L,\nu)\circ (\scr L', \nu')\]
  is the composition $\scr L\circ \scr L'=\scr L\times_{\scr M_2} \scr L'$ in $\mathsf{OSC}$ (if $\scr L$ and $\scr L'$ are composable) and the orientation $\nu\circ\nu'$ of the projection
  \[\scr L\times_{\scr M_2} \scr L' \to \scr L' \to \scr M_3 \]
  is the composition of   the lift of $\nu$ (from $\scr L \to \scr M_2$ to $\scr L\times_{\scr M_2} \scr L' \to \scr L'$)  with $\nu'$ (an orientation of $\scr L' \to \scr M_3$).
\end{blk}

The main related result of \cite{se-qosc} (where we take care of orientations, neglected in \cite{se-qosc}) is
\begin{thm}
  The assignment $\mathsf{Q}\colon \scr M\mapsto\scr M, (\scr L,\nu)\mapsto \delta_{\scr L,\nu},$ is a functor
  \[\mathsf{Q}\colon\mathsf{OOSC}\to\mathsf{QOSC}\]
  in the sense that it sends identity morphisms to identity morphisms and whenever $(\scr L,\nu)\circ(\scr L',\nu')$ is defined then
  \begin{equation}\label{eq:Q-is-functor}
    \mathsf{Q}\bigl((\scr L,\nu)\circ(\scr L',\nu')\bigr) = \mathsf{Q}(\scr L,\nu) \circ \mathsf{Q}(\scr L',\nu').
  \end{equation}
  The functor $\mathsf{Q}$ is (in the obvious way) symmetric monoidal.
\end{thm}

The theorem can be proved as follows. We certainly have
\begin{equation}\label{eq:aux-f}
  \delta_{\scr L, \nu}\circ\delta_{\scr L', \nu'} = f\, \delta_{(\scr L,\nu)\circ(\scr L',\nu')}
\end{equation}
for some $f\in C^\infty(\scr L \circ \scr L')$, simply because we're composing two $\delta$-like integral kernels. By \eqref{eq:Delta-composition} then
\[\Delta(f\, \delta_{(\scr L,\nu)\circ(\scr L',\nu')}) = 0.\] 
In view of \eqref{eq:Delta-f-delta} this implies that $f$ is locally constant.

We need to check that $f=1$. It can be done by considering a special case and then using a deformation argument. For the special case we take
\[ * \xrightarrow{(\Pi\, V^*\!, \nu')} \Pi\, T^*V \xrightarrow{( V, \nu)} * \]
where $V = V_0 \oplus \Pi\, V_1$ is a vector superspace ($V_0$ and $V_1$ are vector spaces) and so $\Pi\, T^*V = V \oplus \Pi\, V^*$. In this case $\nu$ is an orientation of $V_0$ and $\nu'$ a co-orientation of $V_1^*$ in $V_0 \oplus V_1^*$, i.e.\ again an orientation of $V_0$. Checking that
\[ \delta_{V,\nu}\circ \delta_{\Pi V^*,\nu'} =
  \begin{cases}
    \phantom{-}1 & \text{if $\nu=\nu'$}\\
    -1 & \text{if $\nu\neq \nu'$}
  \end{cases}
\]
is immediate. It then follows that if $(\scr L_1, \nu_1)\colon \scr M_1\to *$ and $(\scr L_3,\nu_3)\colon *\to\scr M_3$ then for the two morphisms in $\mathsf{OOSC}$
\begin{equation}\label{eq:LpL-special}
  \scr M_1  \xrightarrow{(\scr L_1, \nu_1)\times(\Pi\, V^*\!, \nu')} \Pi\, T^*V \xrightarrow{( V, \nu)\times (\scr L_3,\nu_3)} \scr M_3
\end{equation}
the identity \eqref{eq:Q-is-functor} is satisfied.

The final bit is the deformation argument. Given composable morphisms in $\mathsf{OOSC}$
\[ \scr M_1  \xrightarrow{(\scr L'\!, \nu')} \scr M_2 \xrightarrow{(\scr L,\nu)} \scr M_3\]
we locally write each $\scr M_i$ as $\Pi T^* V_{(i)}$ in such a way that $\scr L'$ is given by a generating function $g'$ on $V_{(1)} \oplus \Pi\,V_{(2)}^*$ and $\scr L$ by a generating function $g$ on $V_{(2)} \oplus \Pi\,V_{(3)}^*$ (it is (locally) possible because $\scr L$ and $\scr L'$ are composable). If both $g$ and $g'$ are $0$, we are in the case \eqref{eq:LpL-special}. If $g$ and $g'$ are general, we replace them by $\chi g$ and $\chi' g'$, where both $\chi$ and $\chi'$ are 1 on some open set and 0 on another. The corresponding $f$ (see \eqref{eq:aux-f}) is 1 at the places where $\chi$ and $\chi'$ are 0, so it is 1 everywhere, and therefore the function $f$ for the original $\scr L$ and $\scr L'$ is also 1 everywhere.

\medskip

The functor $\mathsf{Q}$ is called the \emph{odd quantization functor}. Its existence is quite remarkable, nothing similar is possible in the even symplectic case. For example let  $\scr M$ be a monoid in $\mathsf{OOSC}$, i.e.\ an odd symplectic groupoid, with the structure Lagrangian relations
\begin{align*}
  \scr U \subset  \scr M \times *& \qquad\text{the unit}\\
  \scr P \subset \scr M \times \overbar {\scr M} \times \overbar {\scr M}& \qquad\text{the product}
\end{align*}
and with orientations $\nu_{\scr U}$ and $\nu_{\scr P}$.
Then $\scr M$ is, via $\mathsf{Q}$, a monoid also in $\mathsf{QOSC}$, with unit $\delta_{\scr U,\nu_{\scr U}}$ and product $\delta_{\scr P,\nu_{\scr P}}$. Therefore the space
\[ A= \on{Hom}(*,\scr M) = \on{Ber}^{1/2}_{-\infty}(\scr M) \]
is a differential $\Z/2\Z$-graded associative `algebra' (with a partially defined product). 

As the simplest example, if $M$ is a manifold then $\Pi\,T^*M$ is naturally a commutative symplectic groupoid; in this case the product on the subspace
\[ \on{Ber}^{1/2}(\Pi\,T^*M) \subset \on{Ber}^{1/2}_{-\infty}(\Pi\,T^*M) = A \]
is well defined. From \S\ref{ssec:odd-Fourier} we know that
\[ \on{Ber}^{1/2}(\Pi\,T^*M) \cong \Pi^n\Omega(M) \]
and indeed the product is just the standard product of differential forms. For non-commutative symplectic groupoids (with non-zero odd Poisson structure induced on $\scr U$) we then get non-commutative versions of differential forms \cite{se-qosc}.

\section{Families of Lagrangian submanifolds}\label{sec:families}

The isomorphism \eqref{eq:hdens-iso-forms} or \eqref{eq:hdens-iso-forms-gr} between differential forms and half-densities plays an important role in BV formalism. If the end result of a calculation is a half-density on $\Pi\, T^*M$ then $M$ is typically seen as a moduli space and the result is interpreted as a differential form on $M$, see \cite{mn-we} for an in-depth tour-de-force example. 

On the other hand, as in any dg-setup, it is useful to consider differential families parametrized by a manifold $\Lambda$, more precisely, to study $\Omega\bigl(\Lambda,\on{Ber}^{1/2}(\scr M)\bigr)$ instead of $\on{Ber}^{1/2}(\scr M)$, with the total differential $d+\Delta$. The fact that we have the isomorphism of complexes
\[ \Omega\bigl(\Lambda,\on{Ber}^{1/2}(\scr M)\bigr) \cong \Pi^{\dim\Lambda}\on{Ber^{1/2}}(\Pi\,T^*\Lambda\times\scr M)\]
allows us to stay just with half-densities and therefore fully within the usual BV formalism.
\medskip

As an example, suppose that we have a proper $\Lambda$-family of immersed oriented Lagrangian submanifolds of an odd symplectic manifold $\scr M$. This means that we have a properly immersed sub-(super)manifold
\[\scr L \subset \Lambda \times \scr M\]
such that the projection $p_\Lambda\colon\scr L\to \Lambda$ is a submersion and such that the fibers of this submersion are Lagrangian (the properness and orientation condition is for the bodies of these supermanifolds).

  A lift 
\[\eta\colon\scr L\to \Pi\,T^*\Lambda \times \overbar{\scr M}\] 
of $\scr L$ is Lagrangian submanifold of $\Pi\,T^*\Lambda \times \overbar{\scr M}$ iff $\eta$, seen as a $1$-form 
\[\eta\in\Gamma(p_\Lambda^*\Pi\, T^*\Lambda)\eqcolon\Omega^{1,0}(\scr L)\subset\Omega^1(\scr L),\]
satisfies $d\eta = p_{\!\scr M}^*\, \omega$. Let $\tilde{\scr L} = \eta(\scr L)\subset \Pi\, T^*\Lambda \times \overbar{\scr M}$ be the resulting Lagrangian submanifold. Then
\[ \delta_{\tilde{\scr L}} \in \on{Hom}(\scr M, \Pi\,T^*\Lambda) = \on{Ber}^{1/2}_{-\infty}(\Pi\, T^*\Lambda \times \overbar{\scr M}, \on{Or}_{\scr M})\]
is $\Delta$-closed. It gives us a morphism of complexes
\[\on{Ber}^{1/2}_c(\scr M) \to \on{Ber}^{1/2}(\Pi\, T^*\Lambda).\]

If we compose this morphism of complexes with the isomorphism $\on{Ber}^{1/2}(\Pi\, T^*\Lambda)\cong\Pi^{\dim\Lambda}\,\Omega(\Lambda)$ from \S\ref{ssec:odd-Fourier}, we get 
\[\on{Ber}^{1/2}_c(\scr M)\to\Pi^{\dim\Lambda}\,\Omega(\Lambda),\quad\sigma\mapsto\la\delta_{\tilde{\scr L}}, \sigma\ra \mapsto \la \delta_{\scr L},e^{-\eta}\sigma\ra \]
(the factor $e^{-\eta}$ comes from the  Fourier transform on the fibres of $\Pi\,T^*\Lambda$).

This morphism of complexes $\on{Ber}^{1/2}_c(\scr M)\to\Pi^{\dim\Lambda}\,\Omega(\Lambda)$, which appeared as an ingenious construction (not using the fact that $\eta$ provides a Lagrangian lift of $\scr L$) in \cite{mi-sch} and which was improved and used in \cite{ge-po}, is thus simply a special case of the calculus of half-densities.

\end{document}